\documentclass[10pt]{bmc_article}    

\usepackage{cite} % Make references as [1-4], not [1,2,3,4]
\usepackage{url}  % Formatting web addresses  
\usepackage{ifthen}  % Conditional 
\usepackage{multicol}   %Columns
\usepackage[latin1]{inputenc} %UNIX support if unicode package fails
\usepackage{comment}
\usepackage{color}
\usepackage{amsmath}
\usepackage{amsthm}
\usepackage{amsfonts}
\usepackage{amsopn}
\usepackage{amssymb} 
\usepackage{graphicx,graphics,epsf,epsfig}
\usepackage{setspace} 

\newboolean{publ}

\newenvironment{bmcformat}{\begin{raggedright}\baselineskip20pt\sloppy\setboolean{publ}{false}}{\end{raggedright}\baselineskip20pt\sloppy}

\begin{document}
\begin{bmcformat}

% Added by authors
%
\onehalfspacing
\parskip 0.2cm
\parindent 1cm


\title{The stochastic behavior of a molecular switching circuit with feedback}
 
%%%%%%%%%%%%%%%%%%%%%%%%%%%%%%%%%%%%%%%%%%%%%%
%%                                          %%
%% Enter the authors here                   %%
%%                                          %%
%% Ensure \and is entered between all but   %%
%% the last two authors. This will be       %%
%% replaced by a comma in the final article %%
%%                                          %%
%% Ensure there are no trailing spaces at   %% 
%% the ends of the lines                    %%     	
%%                                          %%
%%%%%%%%%%%%%%%%%%%%%%%%%%%%%%%%%%%%%%%%%%%%%%

\author{Supriya Krishnamurthy\correspondingauthor$^{1,2}$%
       \email{Supriya Krishnamurthy\correspondingauthor - supriya@sics.se}%
      \and
         Eric Smith\correspondingauthor$^3$%
         \email{Eric Smith\correspondingauthor - desmith@santafe.edu}
      \and
         David Krakauer$^3$%
         \email{David Krakauer - krakauer@santafe.edu}
       and 
         Walter Fontana$^4$%
         \email{Walter Fontana - walter@hms.harvard.edu}%
      }

%%%%%%%%%%%%%%%%%%%%%%%%%%%%%%%%%%%%%%%%%%%%%%
%%                                          %%
%% Enter the authors' addresses here        %%
%%                                          %%
%%%%%%%%%%%%%%%%%%%%%%%%%%%%%%%%%%%%%%%%%%%%%%

\address{%
    \iid(1)Swedish Institute of Computer Science, Box 1263, SE-164 29 Kista, Sweden\\
    \iid(2)Department of Information Technology and Communication, KTH -- Royal Institute of Technology, SE-164 40 Kista, Sweden \\
    \iid(3)Santa Fe Institute, 1399 Hyde Park Road, Santa Fe, NM 87501, USA\\
    \iid(4)Department of Systems Biology, Harvard Medical School, 200 Longwood Avenue, Boston MA 02115 
}%

\vspace*{-2.2cm}

\maketitle

%%%%%%%%%%%%%%%%%%%%%%%%%%%%%%%%%%%%%%%%%%%%%%
%%                                          %%
%% The Abstract begins here                 %%
%%                                          %%
%% The Section headings here are those for  %%
%% a Research article submitted to a        %%
%% BMC-Series journal.                      %%  
%%                                          %%
%% If your article is not of this type,     %%
%% then refer to the Instructions for       %%
%% authors on http://www.biomedcentral.com  %%
%% and change the section headings          %%
%% accordingly.                             %%   
%%                                          %%
%%%%%%%%%%%%%%%%%%%%%%%%%%%%%%%%%%%%%%%%%%%%%%

\begin{abstract}
        % Do not use inserted blank lines (ie \\) until main body of text.
        
        \paragraph*{Background:} Using a statistical physics approach, we study the stochastic switching behavior of a model circuit of multisite phosphorylation and dephosphorylation with feedback. The circuit consists of a kinase and phosphatase acting on multiple sites of a substrate that, contingent on its modification state, catalyzes its own phosphorylation and, in a symmetric scenario, dephosphorylation. The symmetric case is viewed as a cartoon of conflicting feedback that could result from antagonistic pathways impinging on the state of a shared component.
      
        \paragraph*{Results:} Multisite phosphorylation is sufficient for bistable behavior under feedback even when catalysis is linear in substrate concentration, which is the case we consider. We compute the phase diagram, fluctuation spectrum and large-deviation properties related to switch memory within a statistical mechanics framework. Bistability occurs as either a first-order or second-order non-equilibrium phase transition, depending on the network symmetries and the ratio of phosphatase to kinase numbers.  In the second-order case, the circuit never leaves the bistable regime upon increasing the number of substrate molecules at constant kinase to phosphatase ratio. 

        \paragraph*{Conclusions:} The number of substrate molecules is a key parameter controlling both the onset of the bistable regime, fluctuation intensity, and the residence time in a switched state. The relevance of the concept of memory depends on the degree of switch symmetry, as memory presupposes information to be remembered, which is highest for equal residence times in the switched states. 

        \paragraph*{Reviewers:} This article was reviewed by Artem Novozhilov (nominated by Eugene Koonin), Sergei Maslov, and Ned Wingreen. 
\end{abstract}

\ifthenelse{\boolean{publ}}{\begin{multicols}{2}}{}

%%%%%%%%%%%%%%%%%%%%%%%%%%%%%%%%%%%%%%%%%%%%%%
%%                                          %%
%% The Main Body begins here                %%
%%                                          %%
%% The Section headings here are those for  %%
%% a Research article submitted to a        %%
%% BMC-Series journal.                      %%  
%%                                          %%
%% If your article is not of this type,     %%
%% then refer to the instructions for       %%
%% authors on:                              %%
%% http://www.biomedcentral.com/info/authors%%
%% and change the section headings          %%
%% accordingly.                             %% 
%%                                          %%
%% See the Results and Discussion section   %%
%% for details on how to create sub-sections%%
%%                                          %%
%% use \cite{...} to cite references        %%
%%  \cite{koon} and                         %%
%%  \cite{oreg,khar,zvai,xjon,schn,pond}    %%
%%  \nocite{smith,marg,hunn,advi,koha,mouse}%%
%%                                          %%
%%%%%%%%%%%%%%%%%%%%%%%%%%%%%%%%%%%%%%%%%%%%%%

%%%%%%%%%%%%%%%%%%%%%%
%% Open Peer Review %%
%%
\section*{Open peer review}

This article was reviewed by Artem Novozhilov (nominated by Eugene Koonin), Sergei Maslov, and Ned Wingreen. For the full reviews, please go to the Reviewers' comments section.

%%%%%%%%%%%%%%%%
%% Background %%
%%
\section*{Background}
 
Switching plays an important role in cell cycle progression \cite{tyson:01,tyson:02}, cellular differentiation \cite{ferrell:99,ferrell:01,odell:00}, and neuronal memory \cite{lisman:85}. It is therefore a critical property of molecular signal transduction and
gene expression circuits. {\it Ultrasensitivity\/} \cite{goldbeter:81,huang:96} underlies one concept of switching defined by a sharp sigmoidal but continuous response in the concentration of a molecule over a narrow range of a (stationary) signal. A steep ultrasensitive response, however, leads to ``chatter'' when the signal fluctuates across the setpoint \cite{rao:02}. In contrast, {\it bistability\/} \cite{tyson:03} is a form of switching made possible when two stable states, $S_1$ and $S_2$, co-exist over a signal range.  As a consequence, bistable systems exhibit two distinct thresholds as the signal is varied, one at which a transition occurs from $S_1$ to $S_2$ and another at which the system switches back from $S_2$ to $S_1$. The separation of thresholds leads to path dependence or hysteresis, and makes a switched state more impervious to stochastic fluctuations of the signal around the transition point. The difference between bistable and ultrasensitive switching is illustrated in Fig.~\ref{fig:bistable}, where the red curves show switching through bistability and the green curve shows the ultrasensitive form of switching.

Stochasticity in biological processes at the molecular scale has attracted recent attention both experimentally and theoretically \cite{elowitz:02,ozbudak:02,mcadams:99,paulsson:01,thattai:01,rao:02}. 
In this contribution we present a stochastic treatment of bistable switching generated by a reaction network based on a kinase and a phosphatase that phosphorylate and dephosphorylate, respectively, a substrate at multiple sites. The bistability arises from symmetric or asymmetric autocatalytic feedback whereby the substrate catalyzes its own phosphorylation and dephosphorylation (symmetric case), or it catalyzes one but not the other process (asymmetric case); see Fig.~\ref{fig:schemes} for a preview. In statistical physics terminology, the bifurcation from monostable to bistable behavior appears as a non-equilibrium phase transition. Our first step is to derive analytical expressions for the phase diagram (the regions in parameter space where the system exhibits bistable switching) using a mean-field approximation, in which the consequences of correlations among fluctuations are ignored. We then improve on this by taking fluctuations more accurately into account. A major objective is to understand how network structure and molecule numbers affect the thresholds (critical points) of the switch and the longevity of its memory through intrinsic fluctuations.  Some of the analytical results germane to this objective are obtained with a field-theoretic approach whose technical details are laid out in a forthcoming paper \cite{smith:07}. Here we validate these calculations with numerical simulations and discuss their significance. 

There are two possible meanings for the term ``switching" in a bistable system. In one usage the circuit abruptly changes from monostability to bistability (from one to two possible, long-lived macroscopic phosphorylation states), or {\it vice versa\/}, in response to the variation of a parameter. Such a bifurcation in the behavioral repertoire of a system provides a ratchet-like ``checkpoint" \cite{tyson:93}. The other usage of switching refers to the spontaneous fluctuation-induced transition from one macroscopic phosphorylation state to the other within the bistable regime. This corresponds to a cell losing memory of a dynamical state because of noise intrinsic to cellular processes at low molecule numbers. In a stochastic setting, the number of substrate molecules affects both the transition from monostability to bistability and also the residence time in a macroscopic phosphorylation state. Clarifying the role played by fluctuations in these two cases of switching will lead to a better understanding of how reliable, device-like, macroscopic behavior can emerge from low-level stochastic molecular interactions -- a theme dating back at least to von Neumann \cite{JvN}.

In summary, we consider a highly stylized kinetic mechanism of bistability that permits a fairly detailed analytical treatment of its stochastic behavior. We use techniques from non-equilibrium statistical mechanics, which may seem unfamiliar to biologists. Yet, these methods, in conjunction with a model emphasizing a pattern more than biological literalism, convey useful lessons in stochastic reaction-kinetics of biological systems.

\subsection*{Model and biological context}

Molecular signal transduction involves the covalent modification of proteins, giving rise to chemically distinct protein states. A widely occuring modification is phosphorylation, which is the transfer of a phosphate group from an ATP molecule to a tyrosine, serine or threonine residue (referred to as phosphoacceptors) of a target protein by means of a kinase. The 1663 proteins documented in version 3.0 of the Phospho.ELM database \cite{kreegipuu:99} contain between $1$ and $26$ phosphorylatable sites per protein. A ``serine/arginine repetitive matrix 2" protein (SwissProt accession Q9UQ35) exhibits $96$ sites. A list of proteins with $10$ phosphorylatable sites includes, for example, a glutamate-gated ion channel, lamin, glycogen synthase, DNA topoisomerase 2, insulin receptor, and various protein kinases. 

In the system considered here, the target protein is present in a fixed number of copies. Both phosphorylation and dephosphorylation events are catalyzed
concurrently by kinase and phosphatase enzymes whose numbers are also
fixed at the outset. In current terminology \cite{elowitz:02,paulsson:04}, we focus on the intrinsic noise of the reaction system -- the fluctuations in the microscopic phosphorylation states of proteins due to the probabilistic nature of chemical recations at fixed component numbers -- not the additional (extrinsic) noise caused by fluctuations in the numbers of target, kinase, and phosphatase molecules. 

The fully phosphorylated form of a protein often exhibits a distinct catalytic activity, possibly of the kinase or phosphatase type. For example, at each level of the mitogen-activated protein kinase (MAPK) cascade, a signaling circuit widely duplicated and diversified within eukaryotes \cite{kultz:98}, the fully phosphorylated form of the target protein acquires the ability to act as a kinase for the subsequent level. In this paper we will only study a single level or target protein with multiple phosphoaccepting sites.

The fully phosphorylated form of a protein may directly or indirectly promote its own phosphorylation. This is again the case in the MAP kinase cascade, where such feedback reaches across levels. Once fully phosphorylated, the protein at the third level of the cascade catalyzes the accumulation of the kinase at the first level \cite{ferrell:01}, thereby feeding back on its own activation.  A positive feedback with a suitable nonlinearity leads to bistability \cite{ferrell:01}, and the cascade architecture provides that nonlinearity. Bistability can also occur with linear feedback, as in a single phosphorylation / dephosphorylation loop, wherein the phosphorylated substrate catalyzes its own phosphorylation, provided the phosphatase reaction is saturable \cite{lisman:85}. If, on the other hand, all reaction velocities are linear in the substrate, as is the case for large Michaelis constants and small substrate concentrations, the needed nonlinearity can still be generated by multisite phosphorylation. This is the case studied here. We mention for completeness that multisite phosphorylation with saturation kinetics at each modification step can lead to bistability even in the absence of feedback \cite{markevich:04,craciun:06}.

\begin{figure}[!t]
\begin{center} 
\includegraphics[scale=0.7]{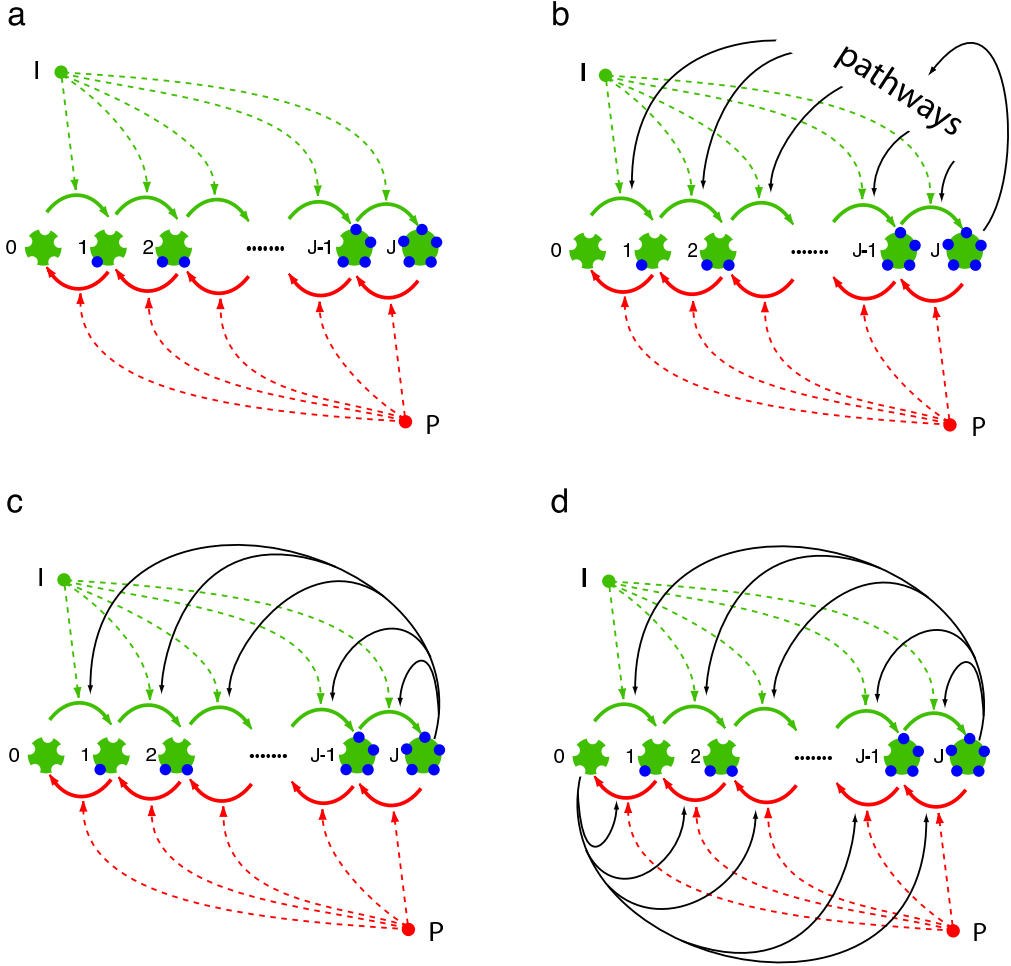} 
\caption{Multisite phosphorylation and feedback structure.
\label{fig:schemes}
}
\end{center}
\end{figure}

We thus consider a protein with $J$ sites that are phosphorylated by a kinase and dephosphorylated by a phosphatase, as illustrated in Fig.~\ref{fig:schemes}a. The numbers of kinase molecules, $I$, and phosphatase molecules, $P$, are fixed. The enzymes are assumed to operate in the linear regime where complex formation is not rate limiting. The site modifications occur in a specific order, thus sidestepping combinatorial complexity. Furthermore, phosphorylation (dephosphorylation) of substrate proteins is assumed to follow a distributive mechanism, whereby a kinase (phosphatase) enzyme dissociates from its substrate between subsequent modification events \cite{ferrell:97,burack:97}. These assumptions lead to $J+1$ states for the substrate. The phosphorylation chain with feedback is shown in the bottom half of Fig.~\ref{fig:schemes}. Panel (c) depicts an asymmetric topology in which the fully phosphorylated substrate catalyzes its own phosphorylation, while panel (d) shows the symmetric version in which a substrate molecule is bifunctional, acting as both a kinase and phosphatase depending on its modification state. Kinase $I$ and phosphatase $P$ are exogenous forces on the modification of the substrate, but the feedback is an endogenous force whose strength is proportional to the occupancy of the end-states of the chain. This occupancy is subject to intrinsic fluctuations and depends on the total number of substrate molecules.

Although each modification step is kinetically linear in both the number of substrate molecules and modification enzymes, the phosphorylation / dephosphorylation chain of Fig.~\ref{fig:schemes}a (no feedback) responds in a nonlinear fashion to a change in the ratio of kinase to phosphatase numbers. Absent feedback, the chain behaves like a random walk with drift and reflecting barriers in the space of substrate phosphoforms. The random walker hops to the right (phosphorylation) with probability $I/(I+P)$ and to the left (desphosphorylation) with probability $P/(I+P)$. As a result of drift there is a geometric multiplier $x$ linking the expected occupancies of phosphoforms. In the absence of feedback, this multiplier is simply $I/P$, the ratio of kinase to phosphatase numbers. For all mean-field treatments, drift causes the expected occupancy of molecules in adjacent phosphorylation states to satisfy
\begin{equation}
\dfrac{\left< n_j\right>}{N}=x\dfrac{\left< n_{j-1}\right>}{N}.
\end{equation}
Thus the expected fractions of fully phosphorylated, $\left< n_J\right>/N$, and unphosphorylated substrate molecules, $\left< n_0\right>/N$, are related as 
\begin{equation}
\dfrac{\left< n_J\right>}{N}=x^J\dfrac{\left< n_0\right>}{N}.
\end{equation}
The geometric progression is a conventional result of detailed balance.
Since by construction $\sum_{j=0}^{J}\left< n_j\right>/N= 1$, we obtain
\begin{equation}
\dfrac{\left< n_J\right>}{N}=\begin{cases}
\dfrac{x-1}{x-x^{-J}}& \text{if $x\ne 1$},\\[0.2cm]
\dfrac{1}{J+1}& \text{if $x=1$.} \label{eq:nofeedback}
\end{cases}
\end{equation}
As the number of sites $J$ tends to infinity, we observe
\begin{equation}
\dfrac{\left< n_J\right>}{N}\xrightarrow[J\rightarrow\infty]{} \begin{cases}
0& \text{if $x\le 1$},\\[0.2cm]
1-\dfrac{1}{x}& \text{if $x>1$}. \label{eq:displ-hyperbola}
\end{cases}
\end{equation}
When the fully posphorylated end-state feeds back on the chain, it alters $x$,  affecting end-state occupancy as described in equation (\ref{eq:nofeedback}). As the end-state occupancy increases, the efficacy of feedback catalysis in moving phosphoforms to the right eventually diminishes as the chain runs out of unphosphorylated substrate. The relevant observation here is that even without the individual reactions exhibiting saturation kinetics, a multisite phosphorylation chain with $J\ge 2$ produces sufficient nonlinearity for feedback to induce bistability. The addition of feedback changes only the functional form of $x$ in the preceding equations, as detailed below. 

\subsection*{Remarks on realism}

We have chosen phosphorylation over other types of post-translational modification for the sake of concreteness. The feedback topology of the model caricatures a few elements present in biological systems. One such element is the competition between antagonistic pathways that may underlie cellular decision processes (for example \cite{gaudet:05}). A multisite phosphorylation chain of the type considered here could function as an evaluation point between competing and antagonistic pathways influenced by different active phosphoforms of the chain, provided these pathways feed back to the chain. In a less extreme case, the fully phosphorylated form activates another kinase which then interacts with the chain. In these scenarios, feedback is mediated by a series of intervening processes, which may well affect the propagation of fluctuations. Yet, if delays are not too large, the collapsed scheme of Fig.~\ref{fig:schemes}c could be a reasonable proxy with the added benefit of mathematically tractability. 

A scenario corresponding more literally to our model involves a bifunctional substrate capable of both kinase and phosphatase activity, depending on the substrate's modification state. One example is the HPr kinase/P-Ser-HPr phosphatase (HprK/P) protein, which operates in the phosphoenolpyruvate:carbohydrate phosphotransferase system of gram-positive bacteria. Upon stimulation by fructose-1,6-bisphosphate, HprK/P catalyzes the phosphorylation of HPr at a seryl residue, while inorganic phosphate stimulates the opposing activity of dephosphorylating the seryl-phosphorylated HPr (P-Ser-HPr) \cite{kravanja:99}. Another example of a bifunctional kinase/phosphatase is the NRII (Nitrogen Regulator II) protein. It phosphorylates and dephosphorylates NRI. NRI and NRII constitute a bacterial two-component signaling system, in which NRII is the ``transmitter" and NRI the ``receiver" that controls gene expression. NRII autophosphorylates at a histidine residue and transfers that phosphoryl group to NRI. The phosphatase activity of NRII is stimulated by the PII signaling protein (which also inhibits the kinase activity). Several other transmitters in bacterial two-component systems seem to possess bifunctional kinase/phosphatase activity \cite{ninfa:91}.

This completes the sketch of the feedback circuit.  We next discuss its statistical properties.

%%%%%%%%%%%%%%%%%%%%%%%%%%%%
%% Results and Discussion %%
%%
\section*{Results and Discussion}

\subsection*{Formal definitions and result overview}

The state diagram of the circuit is shown in Fig.~\ref{fig:schemes}d.
Sites are indexed by $j \in 0 ,\ldots , J$ representing phosphoforms of the target protein.  We describe a population of $N$ target proteins with a vector $n = \left( n_j \right)$, where component $n_j$ is the number of proteins in phosphorylation state $j$ and $\sum_{j=0}^J n_j = N$.  The distribution of phosphoforms in the protein population changes in time because independent phosphorylation and dephosphorylation events cause individual proteins to hop from a state to a neighboring one along the state chain. We assume that both kinase and phosphatase actions occur on a time scale of 1 unit. Phosphorylations (hops from state $j$ to $j+1$) are catalyzed at rate $I +f_J n_J $, where $I$ is the number of kinase molecules and the second term describes the feedback resulting from the fully phosphorylated state $j = J$. The intensity of this feedback is assumed to be proportional to the {\it instantaneous\/} occupation $n_J$.  The feedback is therefore a fluctuating variable, whose strength is set implicitly by the total number of target molecules $N$. Similarly, dephosphorylation transitions (backward hops from site $j$ to $j-1$) occur at rate $P + f_0 n_0$, where $P$ is the number of phosphatase molecules and $n_0$ is the number of target proteins in state $0$. $f_0$ and $f_J$ are measures of the feedback strengths of sites $0$ and $J$, respectively. In all that follows we only deal with the $2$ cases $f_0=f_J=1$, depicted in Fig.~\ref{fig:schemes}d and referred to as symmetric topology, and $f_0=0, f_J=1$, depicted in Fig.~\ref{fig:schemes}c and referred to as asymmetric topology. It is straightforward to generalise the analysis to other cases.

\begin{figure}[!t]
  \begin{center} 
  \includegraphics[scale=0.7]{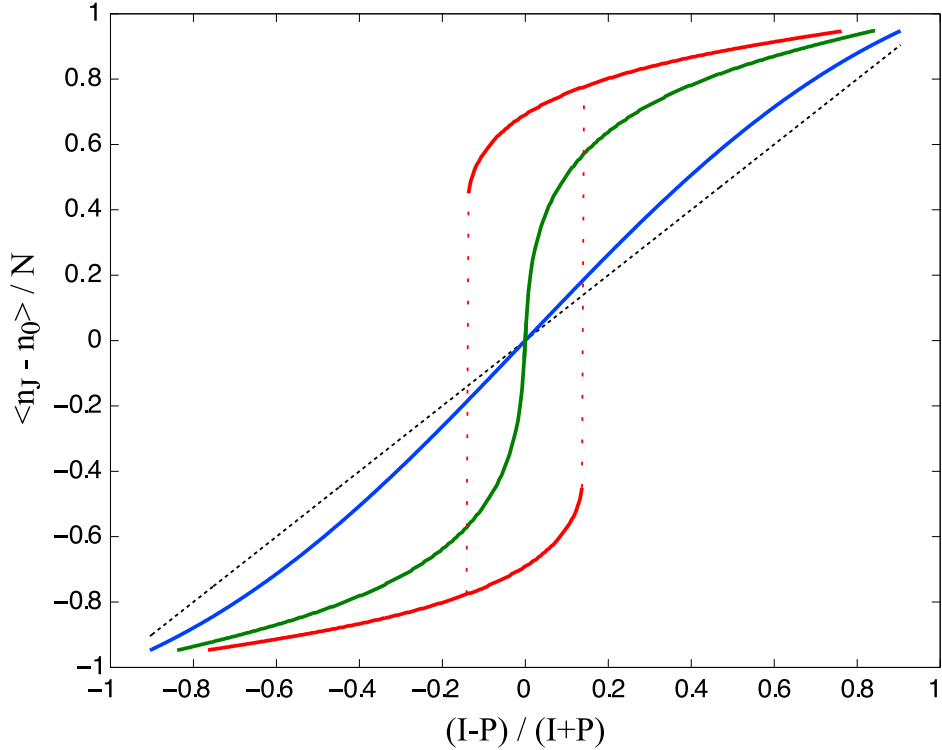}
  \caption{Ultrasensitive and bistable behavior in the symmetric feedback circuit with $J=2$. 
  \label{fig:bistable} 
  }
  \end{center}
\end{figure}

We proceed with an overview of the salient features of this switch as we vary two kinds of asymmetries: parameter asymmetry, $I/P$, and circuit asymmetry. These features are obtained from the analysis detailed in the next section. In a deterministic setting, the transition from a single stable steady-state (monostability) to one with two stable steady-states (bistability) is a saddle-node bifurcation, while in statistical mechanics it appears as a non-equilibrium phase transition. In both symmetric and asymmetric circuits such a phase transition occurs when the ratio of feedback catalysis (the strength of which is set by $N$) to exogenous catalysis (set by $I$ and $P$) is varied.  The number of target molecules $N$ therefore constitutes a degree of freedom for switch control. Simulations, carried out with a simple procedure described in the Methods section, indicate that (for $J\sim 9$ or greater) $N \sim 10$ is enough in practice to start showing a sharp transition in all cases. We find analytically that bistability occurs for any $J\ge 2$, despite the enzymes' operating in the linear regime. 

The case $I = P \equiv q$ for the symmetric circuit has a second-order phase transition between monostability and bistability as the ratio $g \equiv N/q$ is varied, that is, the transition occurs at a critical point at which the order
parameter $\left| \left< n_J - n_0 \right> \right| / N$ changes continuously, see Fig.~\ref{fig:meanfield} (and blue curve in Fig.~\ref{fig:bifurcations}). However, the phase transition is {\it different} for finite $J$ and for infinite $J$. We do not know of any other statistical non-equilibrium model which shows this behavior.  The infinite-$J$ case, although not directly relevant as a biological model, remains important as it constitutes the asymptotic behavior to which all the finite-$J$ solutions converge deep within the bistable regime, where the dynamics of the circuit involves nearly exclusively one boundary (all phosphorylated or all unphosphorylated), so that the configuration space is effectively semi-infinite.  

Parameter asymmetry, $I \neq P$, or circuit asymmetry also generate transitions from monostability to bistabilty; but the transition is now first-order (the order parameter changes discontinuously), see Fig.~\ref{fig:phase_diagram}b and Fig.~\ref{fig:bifurcations}.

For the symmetric circuit and $I=P$, the monostable phase is one in which all the $N$ target molecules are distributed homogenously over the phosphorylation states, while the two degenerate bistable states are those where the population is concentrated on either end of the chain.  The character of the monostable state in cases with any type of asymmetry is different from the symmetric case, in that the $N$ particles follow a density profile peaked around $j=0$ or $j=J$ (depending on the parameters). In Fig.~\ref{fig:phase_diagram} we show the phase diagram for symmetric and asymmetric circuits as a function of parameter asymmetry.

We can solve this model not only for the mean behavior, but also for the fluctuations as well as the residence time in the bistable regime. The latter is of biological interest since it addresses the persistence of switched states (memory) when realized by small numbers of molecules \cite{bialek:01}.

\begin{figure}[!ht]
  \begin{center} 
  \vspace{0.5cm}
  \includegraphics[scale=0.7]{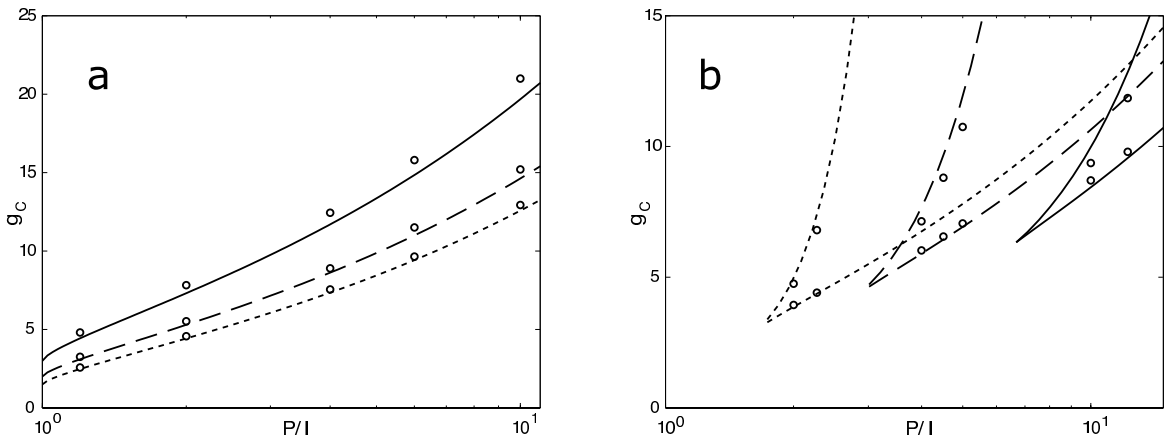}
  \caption{Phase diagrams.
  \label{fig:phase_diagram} 
  }
  \end{center}
\end{figure}

\subsection*{Theoretical analysis}

The system is entirely specified by the probability $\mbox{Pr} \!\left( n \right)$, where $n=(n_0,\ldots,n_J)$ is the vector of occupancies introduced above. The master equation describing the hopping process in the space of phosphoforms for the symmetric circuit (Fig.~\ref{fig:schemes}d) is:
\begin{align}
  \dfrac{d}{dt}\mbox{Pr} \! \left( n \right) =  
  \sum_{j = 0}^{J-1} & \Bigl[ \left(I + n_J - \delta_{J,j+1}\right)
  \left(n_j + 1\right)\mbox{Pr}\!\left( n + 1_j - 1_{j+1} \right) 
  - \left(I + n_J\right) n_j \mbox{Pr} \! \left( n \right)\Bigr.\nonumber \\ 
  & \Bigl. + \left(P + n_0 - \delta_{0,j}\right)\left(n_{j+1} + 1\right)
   \mbox{Pr}\!\left( n - 1_j + 1_{j+1} \right)-\left(P + n_0\right) n_{j+1} \mbox{Pr} \!\left( n \right)\Bigr] . 
\label{eq:master_eqn}
\end{align}

Here $1_j$ represents the vector of zeros with a $1$ in the $j^{\mbox{\scriptsize th}}$ position, and $\delta_{j,j^{'}}=0$ if $j\ne j^{'}$ and $\delta_{j^{'},j^{'}}=1$ (Kronecker delta). In all that follows, we are interested in the stationary state where the left-hand-side of Eq.~\ref{eq:master_eqn} can be set to $0$.

Eq.~(\ref{eq:master_eqn}) is difficult to solve exactly for $J > 1$,
however it is amenable to a mean-field approximation. The idea behind
the mean-field theory is to reduce the system of $N$ target molecules interacting through a fluctuating feedback, to $N$ {\it non-interacting} particles acted on by an {\it averaged} feedback. In this approximation we replace the modification rates $P+ n_0$ and $I+n_J$ in~(\ref{eq:master_eqn}) by effective rates $P+\left< n_0 \right>$ and $I+\left< n_j \right>$, respectively, where the $\left< \mbox{ } \right>$ denotes expectation in the stationary state. This is identical to the spirit of mean-field theory for Ising spin systems where the fluctuating field felt by a spin (on account of its neighbours) is replaced by an average field, which is then determined self-consistently. The reader will find a brief summary of the mean-field approach in the Methods section. 

It is helpful to define a generating function $F(z) \equiv F(z_0,z_1,z_2 \cdots, z_J) \equiv \sum_{\{ n \}} \mbox{Pr} \! \left( n \right)
z_0^{n_0}z_1^{n_1}....z_J^{n_J}$, where the sum is over all
configurations $\{n \}$ that preserve the total number of target molecules.
In the Methods section we solve Eq.~(\ref{eq:master_eqn}) for $\mbox{Pr} \! \left( n \right)$ in the mean-field limit, and obtain the generating function for $N$ substrate molecules as:
\begin{equation}
F(z) = \dfrac{(z_0+xz_1+x^2z_2 + \cdots + x^Jz_j)^N}{(1-x^{J+1})^N}(1-x)^N
\label{eq:genfn}
\end{equation}
where we define
\begin{equation}
x = \left( e^{\lambda/2} + g \left< n_J \right> / N \right) / 
\left( e^{-\lambda/2} + g \left< n_0 \right> / N \right), \; 
g = N / \sqrt{IP}, \text{ and } e^{\lambda} = I/P.
\label{eq:shortcuts}
\end{equation}
The expression for $x$ is just a rewrite of $x=(I+\left< n_J \right>)/(P+\left< n_0 \right>)$ in terms of the coupling strength (control parameter) $g$. The definition of the kinase to phosphatase ratio as an exponential enables convenient use of hyperbolic sines below. Eq.~(\ref{eq:genfn}) obeys the constraints $F(1,1,\cdots,1)=1$ (probability conservation) and ${
\left. \sum_{j=0}^{J} \partial{F}/\partial{z_j} \right| }_{z = 1}=\sum_{j=0}^{J}n_j= N$ (particle number conservation).

From Eq.~\ref{eq:genfn} one obtains $\left< n_j \right> = x \left<
n_{j-1} \right> = x^j \left< n_0 \right>$, and $ N = \left< n_0
\right> \left( 1 - x^{J+1} \right) / \left( 1 - x \right) $. By defining $x = e^{\xi}$, we find
\begin{equation}
\dfrac{\left< n_J - n_0 \right>}{N} = \dfrac{2 \sinh \left( \dfrac{J}{2} \xi \right) \sinh \left( \dfrac{1}{2} \xi \right)}{\sinh \left( \dfrac{J + 1}{2} \xi \right)}.
\label{eqn:ord_parameter_sinh}
\end{equation}
From $\left< n_j \right> = x^j \left< n_0 \right>$ and the definition of $x$, we obtain an expression for $g\left< n_J - n_0 \right>/N$, which, when combined with Eq.~\ref{eqn:ord_parameter_sinh}, enables us to write $g$ in terms of the parameter $\xi$: 
\begin{equation}
g = \dfrac{\sinh\left(\dfrac{\xi-\lambda}{2}\right) \sinh \left(\dfrac{J + 1}{2} \xi \right)}{\sinh\left(\dfrac{\xi}{2}\right)\sinh \left( \dfrac{J - 1}{2} \xi \right)}. 
\label{eqn:g_as_sinh_full}
\end{equation}
In Fig.~\ref{fig:bifurcations} we invert Eq.~\ref{eqn:g_as_sinh_full} to graph $\xi=\log x$ as a function of $g$, showing the transitions to bistability for the symmetric circuit at different values of $\lambda$.

\begin{figure}[!ht]
  \begin{center} 
  \vspace{0.5cm}
  \includegraphics[scale=0.6]{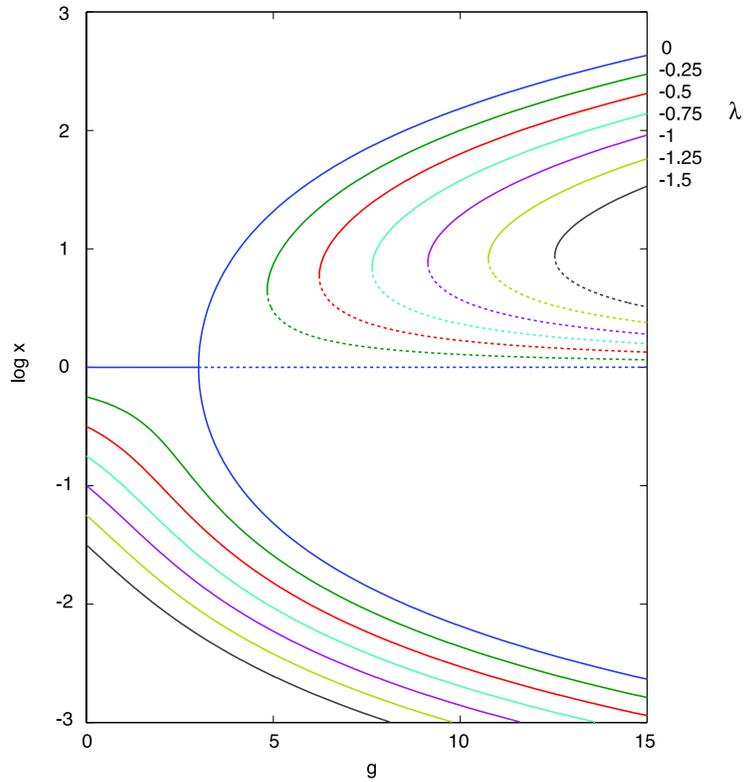}
  \caption{$\xi(g)$ as obtained from inverting Eq.~\ref{eqn:g_as_sinh_full}  for $J=2$. 
  \label{fig:bifurcations} 
  }
  \end{center}
\end{figure}

In all that follows, we consider the simplest case of $I=P$ ($\lambda=0$). Eq.~\ref{eqn:g_as_sinh_full} simplifies to
\begin{equation}
g = \sinh \left(\dfrac{J + 1}{2} \xi \right) / \sinh \left( \dfrac{J - 1}{2} \xi \right).
\label{eqn:g_as_sinh}
\end{equation}
We can now solve for the order parameter $\left|\left< n_J-n_0 \right>\right|/N$ as a function of the parameter $g$ for any $J$. When $I=P$, we find that for $J \ge 2$ there is a critical coupling 
\begin{equation}
g_c = \left( J+1 \right) / \left( J-1 \right),
\end{equation}
such that for $ g \le g_c $ the uniform distribution $ \xi =0 $ with $ n_j / N = 1 / \left( J+1 \right)$ for all $j$ is stable, while for $g > g_c$ the symmetric distribution is a saddle point and the $\xi \neq 0$ solutions are stable. These results are readily extended to the asymmetric circuit, yielding the phase diagrams of Fig.~\ref{fig:phase_diagram}.

\begin{figure}[!ht]
  \begin{center} 
  \vspace{0.5cm}
  \includegraphics[scale=0.6]{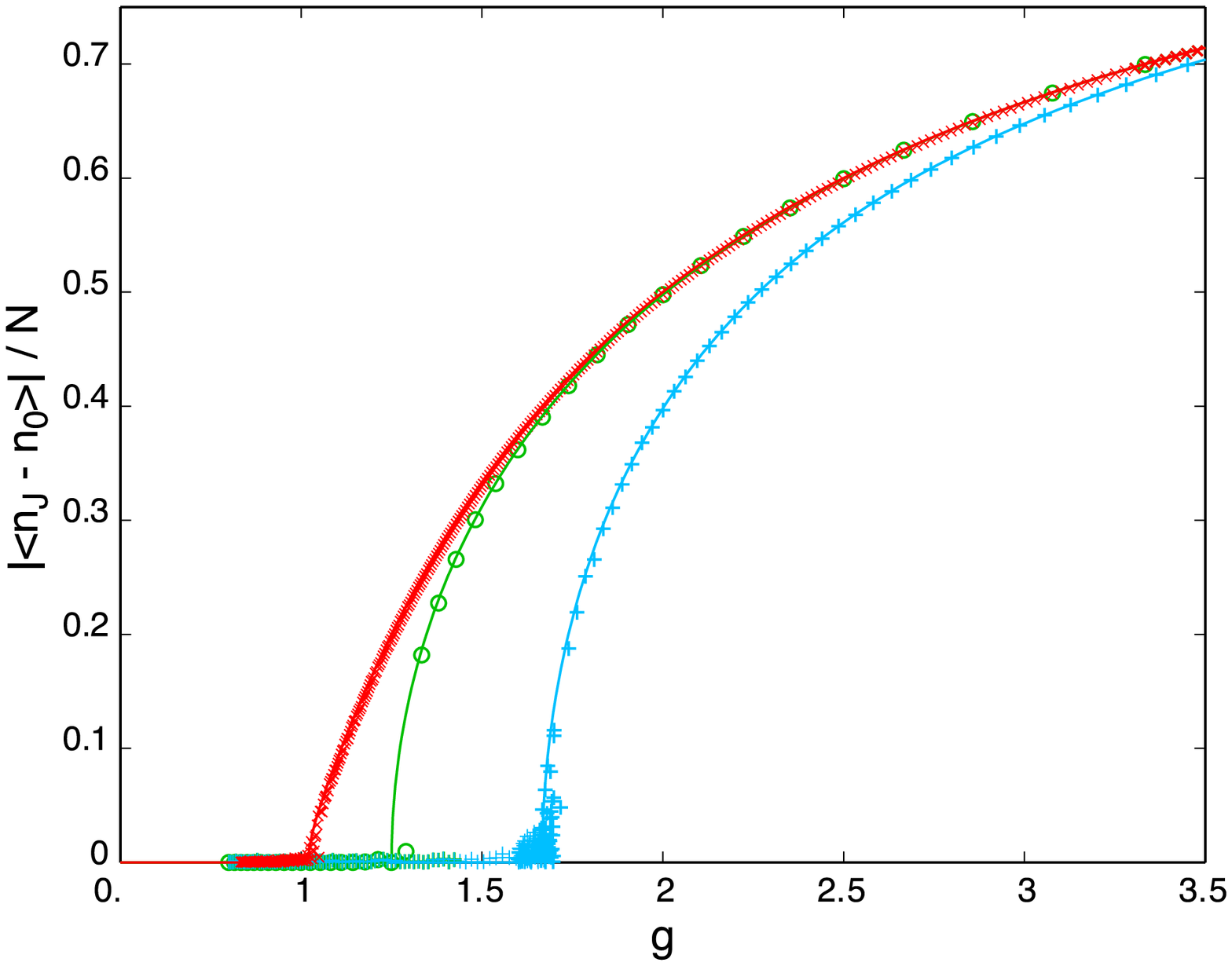}
  \caption{Order parameter $\left| \left< n_J - n_0 \right> \right| / N$ for
   the case $I=P$ in the symmetric circuit. 
  \label{fig:meanfield} 
  }
  \end{center}
\end{figure}

In Fig.~\ref{fig:meanfield} the theoretical estimate of the order parameter from the mean-field approximation is compared against simulations for different values of $J$.  It is seen that the mean-field estimate is very accurate for sufficiently large $N$. The transition at finite $J$ has the character of a Curie-Weiss magnetization transition \cite{goldenfeld:92}, with order parameter scaling as
\begin{equation}
  \dfrac{\left| \left< n_J - n_0 \right>\right|}{N } \approx
  \dfrac{\sqrt{6 J}}{J + 1}{\left(\dfrac{g}{g_c} - 1\right)}^{1/2} . 
\label{eq:ord_p_scaling_CW}
\end{equation}
For any $J$ at $g \gtrsim 1 + 2 \left( g_c - 1 \right)$, the order parameter saturates to a $J$-independent envelope value
\begin{equation}
\dfrac{\left| \left< n_J - n_0\right>\right|}{N} \rightarrow 1 - \dfrac{1}{g} . 
\label{eq:ord_p_scaling_BE}
\end{equation}
Since $g_c - 1 \rightarrow 2 / J$ for large $J$, Eq.~(\ref{eq:ord_p_scaling_BE}) also gives the behavior in the formal $J \rightarrow \infty$ limit.  The derivative of the order parameter converges to 1 in arbitrarily small neighborhoods of the critical point, rather than to $\infty$ as in the Curie-Weiss regime; thus $J \rightarrow \infty$ defines a different universality class than any finite $J$.  Qualitatively, the distinction between large and
small $J$ is determined by whether one or both reflecting boundaries, respectively, are sensed by the near-critical symmetry-broken state.

\begin{figure}[!ht]
  \begin{center} 
  \vspace{0.5cm}
  \includegraphics[scale=0.6]{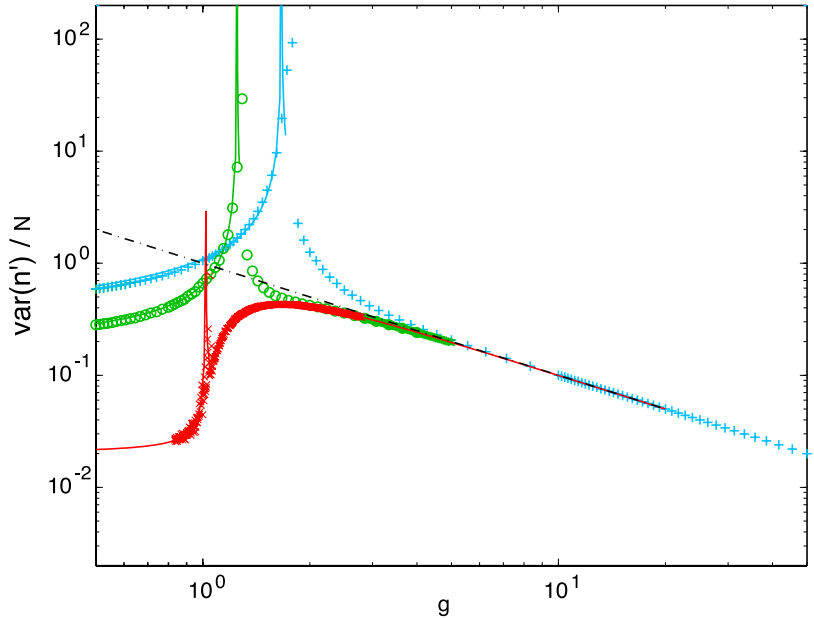}
  \caption{Fluctuations in the order parameter.
  \label{fig:fluctuations} 
  }
  \end{center}
\end{figure}

So far we have described the mean behavior of the system. Eq.~\ref{eq:genfn} also makes predictions for higher moments, in particular the variance of the end-site occupancies $\sigma^2 \left( n' \right) = \left<{\left( n_0 - n_J \right)}^2\right> - {\left< n_0 - n_J\right>}^2$. It is easy to show that for large $g$, $\sigma^2 \left( n'\right) \sim N/g$ (or the noise $\sigma^2/N^2\sim 1/Ng=\sqrt{IP}/N^2$). As shown in Fig.~\ref{fig:fluctuations}, this is a fairly accurate prediction for any $J$. At large $J$, the form $(N/g)(1-g^{-2.5})$ represents an excellent numerical fit for the variance all the way down to the critical $g_c$. 

For a given distance not too far from the critical point, the noise in the order parameter increases with decreasing $J$. At large $g$, far inside the bistable region, the variances for all $J$ converge as indicated above and shown in Fig.~\ref{fig:fluctuations}. Interestingly, for large $J$ the noise increases to a maximum past the critical point (red curve in Fig.~\ref{fig:fluctuations}), while for small $J$ it decreases. At a phenomenological level, we may distinguish between fluctuations and large-deviations. Roughly speaking, a large-deviation state is one that is very unlikely to be reached, but once reached, exhibits a certain duration (residence time). This is the case for the spontaneous switching between stationary states in the bistable regime. A fluctuation does not necessarily have such distinct time scales. At or near criticality, however, there is no clear-cut separation between fluctuations and large-deviations. We return to the topic of large-deviations below.

Fig.~\ref{fig:fluctuations} also demonstrates that the variance {\it diverges\/} at the critical point.  To understand this, we need to go beyond mean-field theory.  We can do this by introducing an operator algebra on a basis of number states \cite{Mattis_Glasser,Cardy}, as done for many reaction-diffusion
systems in physics, and writing the master equation in terms of these number operators. This technique has been recently applied to models of simple gene expression circuits \cite{Sasai_Wolynes}. The technical development of our approach based on the formalism of field-theory and its application to molecular signal transduction is the main subject of a forthcoming manuscript \cite{smith:07}. Here we only sketch the spirit of the approach.

In principle, the master equation \ref{eq:master_eqn} contains the full content of the stochastic process description of our system.  The discrete master equation, however, is a function of the instantaneous occupancies $n_j$ for any particular $j$, and does not lend itself to study of collective changes of phosphorylation state across the whole range of $j$ values.  One purpose for field-theoretic methods is to re-express the content of the master equation in terms of the elementary modes of collective state change. Within this formalism, fluctuations in the phosphorylation state of the circuit can be expanded in a set of basis functions (modes) of the reaction-diffusion operator associated with the master equation. This treatment identifies the lowest mode of the diffusion equation as the collective fluctuation that induces the instability as the coupling $g$ approaches the critical value $g_c$. At the critical point, this mode becomes a linear function of $j$, so that its fluctuations describe an aggregate transformation of the substrate population upward or downward in phosphorylation state, collectively adding or subtracting a linear term to the phosphorylation state distribution. All other modes of diffusion among phosphorylation states decay through the bistability transition, almost as they decay in the monostable region. This form of understanding the mechanism underlying instability in a stochastic circuit is more natural in a field-theoretic representation than in the master equation \ref{eq:master_eqn}. 
  
Within the field-theory formalism, a systematic perturbation theory can be developed about the mean-field limit, yielding expressions for the variance $\sigma^2\left(n' \right)$ plotted in Fig.~\ref{fig:fluctuations} (solid lines). This result confirms the similarity of the finite-$J$ transition to Curie-Weiss ferromagnetism, with a divergence at the critical point scaling as
\begin{equation}
    \dfrac{\left<{\left(n_J - n_0\right)}^2\right> - {\left<n_J - n_0\right>}^2} {2N / \left( J + 1 \right)} \approx\dfrac{\mbox{const} \sim 1}
    {\left( J + 1 \right)\left| g - g_c \right| }
\label{eq:ord_fluct_scaling}
\end{equation}

We can also compute the large-$N$ dependence of the average residence time $\tau$.  This is the average time a state in the bistable regime will last before spontaneous fluctuations cause a transition to the other state.  The residence time is related to the memory (stability) of a stochastic switch.  The persistence of a switched state has been recently studied in other contexts both numerically \cite{warren:04,miller:05} and theoretically \cite{bialek:01,aurell:02,Roma}. We solve the master equation of the system, using the field-theoretic formalism described in \cite{smith:07}, and obtain the leading large-$N$ dependence of the residence time $\tau$ as $\tau \sim e^{N f \! \left( g , J \right)}$, where the function $f \! \left( g , J \right)$ is independent of $N$ at fixed $g$ and $J$. We have numerically evaluated analytical expressions for $N f \!\left( g , J \right)$, obtained from first principles at $J=2$, and find good agreement with Monte-Carlo simulations, as shown in Fig.~\ref{fig:restime}. 

\begin{figure}[!h]
\begin{center}
\vspace{0.5cm}
\includegraphics[scale=0.7]{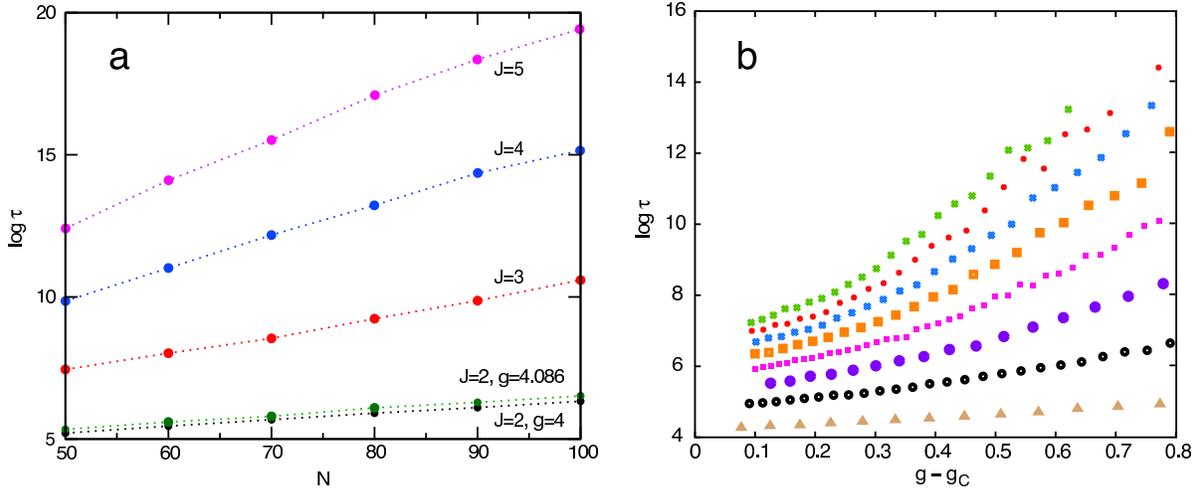} 
\caption{Residence times $\tau$ in the switched state as a function of $J$, $N$, and the distance from the critical point $g-g_c$.
\label{fig:restime}
}
\end{center}
\end{figure}

These simulations were carried out with a rate constant of 1 for each elementary (de)phosphorylation step. To estimate biological time scales, let us assume a catalytic rate constant of phosphorylation, $k_{cat}$, of $0.01$ s$^{-1}$ \cite{markevich:04} (another frequently used value is 1.5 s$^{-1}$ \cite{markevich:04,huang:96}). Dephosphorylation rate constants are roughly an order of magnitude higher \cite{miller:05}. These rate constants refer to the first-order conversion of an enzyme-substrate complex into enzyme and (de)phosphorylated product. Our model, however, does not include complex formation between enzyme and substrate, and (de)phosphorylation reactions are second-order. Because kinase and phosphatase concentrations remain unchanged at each step, we may absorb them into a pseudo first-order rate constant. To estimate that rate constant, we interpret our model as operating in the linear Michaelis-Menten regime, where the pseudo first-order rate constant $k$ becomes $k_{cat}[I]/K_m$, with $K_m$ the Michaelis constant of the enzymatic reaction and $[I]$ the concentration of kinase molecules. Ballpark figures for $K_m$ are $50$ to $500$ nM \cite{markevich:04}. 1 nM corresponds to about $0.6$ particles per femtoliter (fl), or roughly $25$ particles per yeast cell volume ($42$ fl on average). In the simulations and calculations of Fig.~\ref{fig:restime}a we have $I=P=N/g$, to keep the coupling $g$ constant as $N$ changes. The range of the abscissa in Fig.~\ref{fig:restime}a therefore corresponds to kinase concentrations of $0.5$ to $2$ nM, assuming a yeast cell volume. To roughly estimate spontaneous switching times, consider the symmetric circuit with $J=2$, not too far past the critical point, and $N=50$ target particles corresponding to $2$ nM, $I=P=12$ kinase (phosphatase) particles corresponding to $0.5$ nM, and $\tau\sim (1/k)e^{5.2}=181/k$ seconds. Using $k=k_{cat}[I]/K_m=0.01\cdot 0.5/50=10^{-4}$ s$^{-1}$, we obtain an expected residence time of $21$ days in a yeast cell volume. Mammalian cell sizes range anywhere from twice to several hundred times a yeast cell volume, which increases the residence time of a switched state for the same number of particles. The main message is that for the type of symmetric circuit considered here, the spontaneous switching times appear to be in the order of weeks to years for rather moderate numbers of molecules in the minimal circuit. Longer lasting memory may occur in the case of gene regulatory networks \cite{bialek:01}, which typically involve smaller overall rate constants than post-translational signaling events. 

The spontaneous switching time strongly increases with the number of phosphoaccepting sites $J$, as can be seen in Fig.~\ref{fig:restime}. Fits to the numerical data of Fig.~\ref{fig:restime} suggest $\tau\sim\alpha\exp [ N (g-g_c)^2 (\beta+\gamma J^{3/2}) ]$, where $\beta<0$ (slightly), $\gamma>0$, and $\alpha>0$ exhibits a weak $(N,J)$-dependence. It is conceivable that cellular circuits possess switched states spanning a wide range of persistence times and that the evolutionary process acts on the longevity of such states by changing $J$. 

A recent numerical study reports residence times of several decades for a bistable (auto)phosphorylation circuit implicated to function in long term potentiation \cite{miller:05}. The circuit involves calcium/calmodulin-dependent protein kinase II (CaMKII) with $J=6$. Despite an architecture that differs considerably from the circuit considered here, the authors also find an exponential $N$-dependence, predicted in \cite{bialek:01} to be a generic form for domain-switching processes of this type by arguments similar to ours.

%%%%%%%%%%%%%%%%%%%%%%
\section*{Conclusions}

The circuit studied in this contribution represents a situation in
which competing feedbacks (enhancing phosphorylation and, simultaneously, dephosphorylation) impinge on a protein with multiple phosphorylation sites.  The literal reading of the model assumes a bifunctional enzyme, able to switch states (between kinase and phosphatase activity) according to its degree of phosphorylation. On another interpretation the model represents a contraction of indirect feedbacks that are mediated by antagonistic pathways regulated by the target substrate of the circuit. The extent to which this interpretation is appropriate depends on the sensitivities and delays of the intervening pathways. The primary value of this contribution is to present, on a biologically motivated example, a framework (presented in detail elsewhere \cite{smith:07}) for computing the structure, scaling, and magnitudes of fluctuations and residence times for a class of stochastic switching circuits, along with analytical results for the symmetric case.

We show that multisite phosphorylation of a target protein in the linear operating regime of enzymatic catalysis generates bistability when combined with positive feedback. The change from monostability to bistability upon variation of the pertinent parameters can be characterized in terms of statistical mechanics as a non-equilibrium phase transition. The phase structure of the mean values, including the transition between monostable and bistable regions, is well approximated by the classical detailed balance equations of chemical kinetics. Among the parameters controlling system behavior is the number of target (substrate) molecules, not just the numbers of kinase and phosphatase molecules. This is worth emphasizing because the same number strongly affects the persistence times of the switch. Thus, in a molecular circuit like the one considered here, the number of target molecules influences both the existence of the switch and the likelihood that fluctuations cause spontaneous transitions between its stationary states. 

The analytical framework we constructed for the fully stochastic process correctly predicts the phase diagram, fluctuation spectrum, and large deviation properties for a wide range of parameters as obtained from simulations. We summarize a few observations that follow from this analysis.

\begin{enumerate}

\item The behavior of multisite phosphorylation is often characterized in terms of the stationary fraction of fully phosphorylated form only. In this order parameter, multisite phosphorylation without feedback (where switching is of the ``ultrasensitive" form) appears worse than hyperbolic in switching efficiency for any $j>1$, and for $j\to\infty$ the response curve becomes a right-shifted hyperbola \cite{gunawardena:05} (see also Eq.~\ref{eq:displ-hyperbola}), justifying the label ``thresholding device". In Fig.~\ref{fig:bistable}, however, the order parameter is the stationary {\it difference\/} between the occupancies of the fully phosphorylated and unphosphorylated forms. In this rendition, the behavior of multisite phosphorylation increases switching efficiency (as defined in \cite{gunawardena:05}) for any $j>1$ (though not nearly as dramatically as with feedback) and does not exhibit a threshold. These are not contradictory observations, but they illustrate the importance of the choice of observable (order parameter).

\item With the addition of feedback, the stochastic description of the substrate population evidences a non-equilibrium phase transition between monostable and bistable regimes (corresponding to the saddle-node bifurcation in the deterministic picture). In the symmetric case, as is intuitively obvious, the monostable regime has all phosphoforms represented with equal probability, whereas the bistable regime has two degenerate distributions peaked at the  fully phosphorylated or fully dephosphorylated forms. In the fully symmetric case the phase transition is second-order, as in the Curie model of ferromagnetism.  Interestingly, the ferromagnetic analogy breaks down in the limit $J\to\infty$. This limit is relevant in that the finite-$J$ behavior converges to it for increasing feedback at strengths deep inside the bistable regime. Any form of asymmetry considered here, circuit-wise or parameter-wise, leads to a first-order phase transition, as in boiling water.

\item The types of phase transition are helpful in framing questions about switch memory, that is, the system's residence time in the stationary states. Consider the technically imprecise but nonetheless informative metaphor of a switch as a double well potential. At the lower critical point a single well becomes a double well, the minima representing stationary states. Intrinsic fluctuations can knock the system from one well into the other, corresponding to a loss of memory. The concept of memory, however, is meaningful only if both wells have approximately similar depth, because that is when observing the system in either well conveys information. This situation is realized ideally for the symmetric circuit and $I=P$, which is the second-order phase transition. If, on the other hand, the wells are extremely asymmetric, one much deeper than the other, the shallower well is but a rare and short-lived excursion from the deep one. There is little information in observing the system in the deep well, which is to say that there is (almost) nothing for the system to ``remember". The only memory that can be lost (and fast at that) is the occupancy of the shallow well. Our first-principles calculations from \cite{smith:07}, which agree quite well with simulations (see caption to Fig.~\ref{fig:restime}), apply strictly only to the second-order case at low $J$ and near the critical point, but we do not expect them to become meaningless when the transition is weakly first-order. Between the two extremes of second-order and strongly first-order there is an intermediate range for which the mean-field equations might be used to estimate ratios of spontaneous switching times from one well to the other in terms of the inverse ratio of residence times in the respective wells, as required by detailed balance.

\item The number of substrate molecules plays an important control role in selecting the monostable or bistable regimes. We point out that the symmetric circuit has a bistable regime at constant coupling $g$ as $I/P$ is varied (see Fig.~\ref{fig:bistable}), but it remains forever bistable as $N$ is increased past $g_c$ at constant $I/P$ (see Fig.~\ref{fig:phase_diagram}a). Increasing the number of substrate molecules, increases the strength of feedback and reaction rates, thereby favoring bistability (and reducing fluctuations).

\item The magnitude of fluctuations and residence times in a stationary state as a function of the coupling strength $g$ strongly depend on the number of phosphoaccepting sites $J$. The longevity of state memory could thus be tuned by evolving the number of modifiable sites in the target protein of the circuit (in addition to changing rate constants). The present state of knowledge does not permit to assess whether the latter, more structural degree of freedom is actually exploited in the evolutionary tuning of molecular switch circuitry. One would need to identify proteins that are involved in switches (i.e. that exhibit bistable phosphoform distributions) and relate the longevity of their switched state to the number of phosphorylatable sites. Likewise, we do not know the extent to which cellular decision processes hinge on a wide spectrum of memory time scales in post-translational signaling, but such a spectrum seems not to be difficult to achieve. 

\end{enumerate}

%%%%%%%%%%%%%%%%%%
\section*{Methods}

\subsection*{Monte Carlo simulations}

The Monte Carlo simulations are carried out in a simple manner that follows the process described by the master equation \ref{eq:master_eqn}. A lattice of $J+1$ sites (representing states or phosphoforms) is populated with $N$ target molecules (particles) randomly distributed over the lattice. The particles are numbered and we keep track of their location. To simulate the dynamics, particles are moved to adjacent sites on the lattice at discrete and equally spaced time steps. At any one step, each particle has a probability $(q+n_J)/(2q+N)$ of moving to the right and $(q+n_0)/(2q+N)$ of moving to the left, $q\equiv I=P$. If the particle is located at the first (or the last) site, it can only move to the right (or the left) with probability $(q+n_J)/(2q+N)$ (or $(q+n_0)/(2q+N)$, respectively). The system is evolved until it attains steady-state, when all the measurements are made. The normalisation factor $2q+N$ is an inverse time unit needed to make the rates in Eq.~\ref{eq:master_eqn} into probabilities.

\subsection*{The definition of mean-field approximations}

The general form of master equations such as Eq.~(\ref{eq:master_eqn}) may be written
\begin{equation}
    \dfrac{d}{dt}\mbox{Pr}_n =\sum_{n^{\prime}}T_{n, n^{\prime}}\mbox{Pr}_{n^{\prime}}. 
\label{eq:ME_abstracted}
\end{equation}
In Eq.~(\ref{eq:ME_abstracted}) the discrete argument $n$ has been written
as a subscript denoting a vector index rather than a functional argument,
and the sum over contributing terms is written as a matrix transformation
on the vector with components $\mbox{Pr}_n$.  The matrix $T$ is called the {\em
transfer matrix} of the probability process under which $\mbox{Pr}_n$ evolves.

Moments of $\mbox{Pr}$, such as the mean number, are defined through expectations
\begin{equation}
    \left< n \right> = \sum_n n \mbox{Pr}_n .  
\label{eq:n_mean_def}
\end{equation}
Through the master equation~(\ref{eq:ME_abstracted}) the time evolution of
$\left< n \right>$ may be calculated as
\begin{equation}
    \dfrac{d \left< n \right>}{dt} =\sum_n n \dfrac{d \mbox{Pr}_n}{dt} =
    \sum_n n \sum_{n^{\prime}}T_{n , n^{\prime}}\mbox{Pr}_{n^{\prime}}
\label{eq:n_evol_through_P}
\end{equation}
The steady-state solution $d \left< n \right> / dt = 0$ may be solved
through Eq.~(\ref{eq:n_evol_through_P}) as a balance equation between
moments of $\mbox{Pr}$ expressed on the right-hand side of the equality.  In
general, the moment equations are not closed, either for steady or
non-steady states.  The time dependence or steady-state condition for
$\left< n \right>$ is a function of quadratic moments in $n$, whose time
dependence or steady state conditions are in turn functions of higher-order
moments.

The mean-field approximation truncates this hierarchy of dependencies into
a closed system of equations for the expectations $\left< n \right>$.  In
particular, because the steady-state solution to
Eq.~(\ref{eq:n_evol_through_P}), for the master equation~\ref{eq:master_eqn} involves only quadratic moments, the relation can be closed by replacing $I + n_J$ by $\left< I + n_J \right>$ and $P + n_0$ by $\left< P + n_0\right>$ in
Eq.~\ref{eq:master_eqn} and then evaluating the relation~(\ref{eq:n_evol_through_P}) at $d \left< n \right> / dt = 0$.  The net result is to replace terms such as $\left< \left( I + n_J \right) n_j \right>$ by the approximations $\left< I + n_J \right> \left< n_j \right>$.  The resulting feedback operates only through its mean strength $\left< I + n_J \right>=I+\left< n_J \right>$, which motivates the term ``mean field approximation''.

The mean-field approximation of Eq.~(\ref{eq:master_eqn}) leads to the classical equations of detailed balance with hopping rates expressed in terms of mean occupation numbers.  Thus, the mean particle numbers in adjacent
phosphorylation states satisfy the relation
\begin{equation}
   \left< n_{j+1} \right> = x\left< n_j \right> , 
\label{eq:jp1_j_x_reln}
\end{equation}
with the geometric factor
\begin{equation}
  x = \dfrac{I + \left< n_J \right>}{P + \left< n_0 \right>} = 
  \dfrac{e^{\lambda/2} + g \left< n_J \right> / N}{e^{-\lambda/2} + 
  g \left< n_0 \right> / N} , 
\label{eq:x_def_gen_symtop}
\end{equation}
by the definitions, Eq.~(\ref{eq:shortcuts}), in the main text.  

\subsection*{The generating function for the steady-state master equation}

To derive the equilibrium solution for the generating function from
the master equation~(\ref{eq:master_eqn}), first recall the definition
$F(z) \equiv \sum_{\{ n \}} \prod_{i=1}^J z_i^{n_i} \mbox{Pr} \!
\left( n \right)$.  By multiplying Eq.~(\ref{eq:master_eqn}) by
$\prod_{i=1}^J z_i^{n_i}$ and summing over $\left\{ n \right\}$, we
obtain the equivalent time-evolution equation for the generating
function
\begin{align}
  \dfrac{d}{dt} F \! \left( z \right) =  
  \sum_{\left\{ n \right\}}
  \sum_{j = 0}^{J-1} & 
    \Bigl[ \left(I + n_J - \delta_{J,j+1}\right)
     \left(n_j + 1\right)
     \prod_{i=1}^J z_i^{n_i} \mbox{Pr}\!\left( n + 1_j - 1_{j+1} \right) 
   - \left(I + n_J\right) n_j 
     \prod_{i=1}^J z_i^{n_i} \mbox{Pr} \! \left( n \right)
  \Bigr.
\nonumber \\ 
  & \Bigl. + \left(P + n_0 - \delta_{0,j}\right)
    \left(n_{j+1} + 1\right)
    \prod_{i=1}^J z_i^{n_i} \mbox{Pr}\!\left( n - 1_j + 1_{j+1} \right)
  - \left(P + n_0\right) n_{j+1}
    \prod_{i=1}^J z_i^{n_i} \mbox{Pr} \!\left( n \right)\Bigr] . 
\label{eq:master_eqn_GF_raw}
\end{align}
Because $\left\{ n \right\}$ is now a variable of summation, it can be
shifted in the first term of each line of Eq.~(\ref{eq:master_eqn_GF_raw}), to obtain an equation without offsets in the indices,
\begin{equation}
  \dfrac{d}{dt} F \! \left( z \right) =  
  \sum_{\left\{ n \right\}}
  \sum_{j = 0}^{J-1} 
    \left[ \left(I + n_J \right)
      \left( \dfrac{z_{j+1}}{z_j} - 1 \right) n_j
    - \left(P + n_0\right) 
      \left( \dfrac{z_j}{z_{j+1}} - 1 \right) n_{j+1}
    \right] 
    \prod_{i=1}^J z_i^{n_i} \mbox{Pr} \!\left( n \right) . 
\label{eq:master_eqn_GF_shift}
\end{equation}
Since $ \dfrac{\partial}{\partial z_j} \prod_{i=1}^J z_i^{n_i} = 
\dfrac{n_j}{z_j} \prod_{i=1}^J z_i^{n_i}$, and likewise for $j+1$, we
may further condense Eq.~(\ref{eq:master_eqn_GF_shift}) into 
\begin{equation}
  \dfrac{d}{dt} F \! \left( z \right) =  
  - \sum_{\left\{ n \right\}}
    \sum_{j = 0}^{J-1} 
    \left( z_{j+1} - z_j \right)
    \left[ 
      \left(P + n_0\right) 
      \dfrac{\partial}{\partial z_{j+1}} - 
      \left(I + n_J \right)
      \dfrac{\partial}{\partial z_j} 
    \right] 
    \prod_{i=1}^J z_i^{n_i} \mbox{Pr} \!\left( n \right) . 
\label{eq:master_eqn_GF_condense}
\end{equation}

The mean-field approximation consists in replacing the terms $P + n_0$
and $I + n_J$ by their average values in the configurations --
self-consistently determined -- which dominate the
sum~(\ref{eq:master_eqn_GF_condense}).  These are denoted $P + \left<
n_0 \right>$ and $I + \left< n_J \right>$.  Note that with this
approximation, the terms involving $z$ and $\partial / \partial z$ are
separated from the index sum over $\prod_{i=1}^J z_i^{n_i} \mbox{Pr}
\!\left( n \right)$, which may be performed to recover a differential
equation for the scalar function $F \! \left( z \right)$.  Using the
definitions from Eq.~(\ref{eq:shortcuts}) in the main text, and removing an overall scale for time to the front, we then recast
Eq.~(\ref{eq:master_eqn_GF_condense}) as
\begin{equation}
  \dfrac{d}{dt}F \! \left( z \right) =  
  - \dfrac{N}{g}
    \left(
      e^{-\lambda/2} + 
      g \dfrac{\left< n_0 \right>}{N} 
    \right) 
    \sum_{j = 0}^{J-1} 
    \left( z_{j+1} - z_j \right)
    \left( 
      \dfrac{\partial}{\partial z_{j+1}} - 
      x
      \dfrac{\partial}{\partial z_j} 
    \right) 
    F \! \left( z \right) . 
\label{eq:master_eqn_GF_scalar}
\end{equation}
The negative sign reflects the convergence of a finite Markov process
to its stationary distribution. 

The polynomials $\left( z_{j+1} - z_j \right)$ in
Eq.~(\ref{eq:master_eqn_GF_scalar}) are not needed to identify a
solution to $dF / dt = 0$, as any power of the variable 
\begin{equation}
  \zeta \! \left( z \right) \equiv 
  \sum_{i=0}^J x^i z_i , 
\label{eq:zeta_def}
\end{equation}
is annihilated by the derivative terms in parentheses in
Eq.~(\ref{eq:master_eqn_GF_scalar}), independently at each $j$.  The
particular combination
\begin{equation}
  F_1 \! \left( z \right)\equiv 
  \dfrac{
    \zeta \! \left( z \right)
  }{
    \sum_{i=0}^J x^i
  } = 
  \dfrac{
    \zeta \! \left( z \right) \left( 1 - x \right)
  }{
    \left( 1 - x^{J+1} \right)
  }
\label{eq:zeta_normed_def}
\end{equation}
satisfies the normalization condition $F_1 \! \left( z \equiv 1
\right) = 1$ for a generating function, but its first-derivative
satisfies ${ \left. \sum_{j=0}^J \partial / \partial z_j F_1 \right|
}_{z \equiv 1} = 1$, identifying $F_1$ as the generating function for
a single particle.  For $N$ particles, the generating function $F
\! \left( z \right) = F_1^N$, which is a general relation for the
non-interacting particles described by the mean-field approximation,
and which recovers Eq.~(\ref{eq:genfn}) in the main text.

%%%%%%%%%%%%%%%%%%%%%%%%%%%%%%%%
\section*{Authors contributions}

S.K. and E.S. contributed equally to this work and are joint first authors. S.K and E.S developed the field-theoretic formalism and carried out the analytical and numerical calculations. W.F. and D.K. suggested the problem, interpreted results in a biological light and, jointly with S.K and E.S., wrote the manuscript.

%%%%%%%%%%%%%%%%%%%%%%%%%%%%%%%%
\section*{Reviewers' comments}
\subsection*{Reviewer's report 1}
{\it Ned Wingreen, Department of Molecular Biology, Princeton University\/}

\noindent This paper reports a careful mathematical analysis of a model biochemical circuit with feedback. The model consists of a protein with multiple phosphorylation sites, which can feed back to positively affect its own phosphorylation or dephosphorylation rates, but only when fully phosphorylated or dephosphorylated, respectively. An asymmetric model containing only phosphorylation feedback is also considered. The authors show that these simple models display non-equilibrium phase transitions between a monostable and a bistable regime, and show that there is good agreement between mean-field theory results and exact simulations. Importantly, the rates of phosphorylation and dephosphorylation are taken to be linear functions of enzyme density. The bistability emerges instead from the assumption that full phosphorylation (or dephosphorylation) of multiple sites is required for feed back.

\noindent Overall, the paper makes the case that even very simple biochemical circuits can display bistability, with potentially very slow switching between distinct states. An important point raised in the paper is that the concentration of proteins is an essential parameter in determining the regime of behavior. With increasing protein concentration, the relative strength of feedback becomes stronger, and the system is driven into the regime of bistability. This could be important in biological systems where protein concentrations can be actively controlled, e.g. by transcriptional regulation or by regulated protein degradation.

\noindent In the model section, the distinction between ``ultrasensitive" and ``threshold-like" behavior is not immediately obvious. I recommend a figure showing showing examples of these two types of behavior.

\noindent \underline{Author response:} {\it We have added what is now Figure 2 to illustrate both forms of switching, but also to convey a wealth of further information to which we refer at various points in the manuscript.\/}

\noindent From a biological point of view, it would be helpful to relate the switching time estimate to real biological parameters. Even a simple example that ends up with a estimated switching time would be helpful. 

\noindent \underline{Author response:} {\it We have added an extensive paragraph discussing and estimating switching times.\/}

\subsection*{Reviewer's report 2}
{\it Sergei Maslov, Brookhaven National Laboratories\/}

\noindent The manuscript contains an exhaustive analytical and numerical study of the stochastic dynamics of a multistep phosphorylation/dephosphorylation of a single substrate in the presence of a positive feedback. From the mathematical standpoint the results are as complete as they could ever be: even the multi-variable generating function was computed (albeit in the mean-field limit). Expectedly, under favorable conditions the positive feedback generates bistability which is also analyzed by the authors in great detail.

\noindent However, in my opinion authors should make a better case for the biological meaningfulness of these results. 

\noindent Is there any evidence that the (possibly indirect) feedback in real-life systems of this type has a positive sign? Could authors name at least one concrete example where bistability of phosphorylation/dephosphorylation of a single substrate is at least suspected?

\noindent \underline{Author response:} {\it We have added a further example of a protein with bifunctional  kinase / phosphatase activity. It is not known, however, whether these systems have switch character. We point to reference \cite{gaudet:05} as a recent example suggesting competition between antagonistic pathways in cellular decision processes. Again, it is unclear whether such antagonism converges on a shared circuit causing bistability along the lines we consider in our paper.\/}

\noindent Since the case of infinite $J$ has no biological meaning I would recommend removing the discussion of its peculiar mathematical properties from the manuscript.

\noindent \underline{Author response:} {\it We disagree. We now explain in the text (as is clear from the pertinent figure) that the infinite-$J$ behavior is the asymptote for all finite-$J$ cases deep inside the bistable regime. (The finite-$J$ curves converge to the infinite-$J$ curve at large $g$.) It is, therefore, of interest to understand the infinite-$J$ case.\/}

\noindent What are the advantages and disadvantages of different values of J with respect to signal to noise ratio. In other words, are there any biological lessons to be learned from Figure 6?

\noindent \underline{Author response:} {\it We have added a brief paragraph in the pertinent section of the manuscript to verbally point to the fact shown in Figure 6 that noise decreases with $J$ at a given distance from the transition. We also point out that for a given $J$ noise increases past the transition when $J$ is large and decreases when $J$ is small. We don't know whether this has biological significance.\/}

\noindent Since the ``average residence time" $\tau$ is a very important characteristic of any bistable switch used as a memory device could authors attempt to estimate it for some biologically realistic microscopic parameters? For example, in Figure 7 for, say, $N=100$ molecules would $\tau$ be of order of minutes, seconds, milliseconds?

\noindent \underline{Author response:} {\it We have added an extensive paragraph addressing this issue (which was also raised by Dr.~Wingreen).\/}

\subsection*{Reviewer's report 3}
{\it Artem Novozhilov, National Center for Biotechnology Information (nominated by Eugene V Koonin, National Center for Biotechnology Information, NIH)\/}

\noindent The authors formulate and analyze a stochastic model of a molecular switch frequently found in the molecular control circuitry of cells. This model circuit has been a subject of extensive research during the last decade. Within the framework of deterministic models, it was shown that similar models can exhibit bistability and hysteresis, which might be responsible for some of all-or-none  irreversible processes in the cells. The main novelty introduced by the authors is taking into account intrinsic stochasticity of the (de)phosporylation processes and analytical analysis of the resulting stochastic mathematical model (which is in remarkable agreement with the simulation results). Applying formalism from the theory of non-equilibrium phase transitions the authors obtain phase diagram (stratification of the parameter space into mono- or bistable domains) for arbitrary parameters and fluctuation spectrum for the symmetric scheme with equal rates of kinase and phosphatase activity. This is an important contribution because stochastic consideration not only allow one to obtain the phase diagram in terms of mean system behavior, but also to estimate such probabilistic characteristics as the variance of the order parameter and mean residence time of the substrate population in one of the two possible states, thus providing some insight how macroscopic behavior can emerge from stochastic molecular interactions.

\noindent However, I have some comments and questions concerning the text.

\noindent I think that some of the statements presented in the Conclusion section do not really follow from the results and discussion in the main text.  The authors claim that ``The primary value of this contribution is to present a thorough and rigorous stochastic analysis of a switching circuit" is disputable.  

\noindent (a): The derivation of the mathematical results in the case of the mean-field analysis is, in my opinion, quite fragmentary and could be supplemented with a
Mathematical Appendix.

\noindent (b): The results for fluctuation in the order parameter and for residence time are presented without any derivation with \cite{smith:07} to unpublished work. These results also could be included in Mathematical Appendix. 

\noindent (c): The results for fluctuation in the order parameter and for residence time are presented and discussed only in the case of the symmetric feedback and $P=I$ (loosely speaking, not the most probable case). Is it possible to obtain analytical estimates of these characteristics in the general case?  

\noindent (d): Thus, I presume that it would be more appropriate to state that the primary value of the author's contribution is to illustrate, on a biologically relevant example, the analytical framework presented elsewhere (\cite{smith:07}) and show that in some particular cases it is possible to calculate fluctuations and residence times. 

\noindent (e): I, hence, would also question conclusion 6 (here, the way it is stated in the main text, equal rates of(de)phosphorylation are not enough for a second-order phase transition, one should also have the symmetric feedback) and $\ldots$

\noindent (f): $\ldots$ conclusion 7, which, being quite a general assertion on the average residence times, does not follow from the results presented only for the symmetric scheme and $P=I$.

\noindent \underline{Author response:} {\it (a): We have added two detailed appendices in regard. One appendix explains the idea of mean-field theory. The other derives in detail the generating function $F(z)$.

\noindent (b): We understand the frustration of references to unpublished work, but we feel we cannot do justice to the technical details of the field-theory formalism by relegating them into an appendix. The appendix would become a manuscript in its own right. A preprint of the paper will be available shortly and we hope we can link to it from the web-published version of the present article. We believe that the presently revised and extended version of this manuscript decreases its dependency on the forthcoming paper to an acceptable level. 

\noindent (c): We are at the moment unable to provide analytical calculations for residence times in the asymmetric case. In point 3 of the Conclusions we now discuss issues related to residence times in the asymmetric case more extensively. 

\noindent (d): We agree with the reviewer's assessment regarding the primary value of our contribution to the point of paraphrasing his comment in our Conclusions. 

\noindent (e): The second-order phase transition hinges on parameter and circuit symmetry. We are now more explicit throughout the manuscript about the symmetries we refer to. 

\noindent (f): We have reworded what was conclusion 7.
\/}

\noindent The authors could elaborate on the hysteresis (or path dependence) phenomenon (of course, in terms of the order parameter) which can be seen from Figure 4b, where two critical couplings are found. In connection with this, it could be helpful to produce figures similar to Fig. 5 but for $P \ne  I$ and both symmetric and asymmetric feedbacks. This additional plot(s) can also clarify for a biologist the difference between first- and second- order phase transitions. I also suggest to redraw the plot in Fig. 5 such that both branches of the order parameter are shown when $g> g c$ , thus illustrating the concept of bistability.

\noindent \underline{Author response:} {\it The newly added Figure 2 addresses some of these requests and illustrates the concept of bistability. We have added wording to the caption of Figure 4 (new numbering) to remind the reader that the 
modulus folds both symmetric branches into one. From the viewpoint of 
hysteresis, asymmetries affect the shape details of the curves but don't 
impact any essential aspects of the message conveyed by Figure 2. We have 
added what is now Figure 4 to show the bistability transitions for the 
symmetric circuit with asymmetric $I/P$. This illustrates, as the reviewer requested with biologists in mind, the difference between first and second-order transitions. The Figure complements both Figure 3a (phase portrait) and Figure 5 (which is the $I=P$ case compared against simulations and for varying $J$). We have not added a similar figure for the asymmetric circuit, as the phase portrait, Figure 3b, suffices. The addition of Figure 4 has necessitated a few additional equations in the theoretical analysis section.\/}

\noindent The part of the paper dedicated to the average residence time is too short considering its importance. It would be very interesting to see residence times in other cases (e.g., the symmetric feedback and $P \ne I$ ).

\noindent \underline{Author response:} {\it We have now addressed some of these issues, as indicated  in our responses to similar points raised by other reviewers.\/}

\noindent It is indicated in the text (the last paragraph of the Theoretical Analysis section) that ``function $f (g ,J)$ is independent of $N$" but parameter $g$, as defined in the text, depends on $N$ . Can you clarify this point?

\noindent \underline{Author response:} {\it To meaningfully compare residence times for different values of $N$ at fixed $J$ we must also fix the value of $g$, which is done by adjusting $\sqrt{IP}$ to compensate for the fact that $N$ affects the coupling, $g=N/\sqrt{IP}$. The phrase ``$f(g,J)$ is independent of $N$ at fixed $g$ and $J$" expresses exactly this, but we have unpacked it in the 
caption to Figure 7 (new numbering).\/}

\noindent Figure 7 gives an impression that the mean residence time depends only on $N$ and $J$, and is independent of kinase activity, phosphatase activity, and the type of the feedback. Is it really so? 

\noindent \underline{Author response:} {\it See our comments to the previous point.\/}

%%%%%%%%%%%%%%%%%%%%%%%%%%%
\section*{Acknowledgements}
  \ifthenelse{\boolean{publ}}{\small}{}
  
S.K. was supported by the Swedish Research Council. E.S. is grateful to acknowledge Insight Venture Partners for support. The authors wish to thank the reviewers for constructive comments and Jeremy Gunawardena for fruitful discussions.

%%  Bmc_article.bst  will be used to                       %%
%%  create a .BBL file for submission, which includes      %%
%%  XML structured for BMC.                                %%

{\ifthenelse{\boolean{publ}}{\footnotesize}{\small}
 \bibliographystyle{bmc_article}  % Style BST file
  \bibliography{switch} }     % Bibliography file (usually '*.bib' ) 

%%%%%%%%%%%

\ifthenelse{\boolean{publ}}{\end{multicols}}{}

%%%%%%%%%%%%%%%%%%%%%%%%%%%%%%%%%%%
%%                               %%
%% Figures                       %%
%%                               %%
%% NB: this is for captions and  %%
%% Titles. All graphics must be  %%
%% submitted separately and NOT  %%
%% included in the Tex document  %%
%%                               %%
%%%%%%%%%%%%%%%%%%%%%%%%%%%%%%%%%%%

%%
%% Do not use \listoffigures as most will included as separate files

\section*{Figures}

\subsection*{Figure 1 - Multisite phosphorylation and feedback structure.}
  
(a) This panel depicts the basic phosphorylation chain without feedback in which a target protein with $J$ sites is phosphorylated by a kinase $I$ and dephosphorylated by a phosphatase $P$. The ordered succession of phosphorylations yields $J+1$ modification states, labelled $0$, $1$, $\ldots$, $J$. (b) The fully phosphorylated target protein relays a signal into pathways that eventually feed back on the phosphorylation chain. (c) Simplification of (b) in which the fully phosphorylated target protein acquires kinase activity and directly feeds back on the chain. We refer to this network configuration as the asymmetric circuit. (d) Schematic of the network with symmetric feedback in which the substrate protein is bifunctional, whereby the fully (de)phosphorylated form catalyzes (de)phosphorylation of its own precursors. 

\subsection*{Figure 2 - Ultrasensitive and bistable behavior in the symmetric feedback circuit with $J=2$.}
  
The ordinate is the order-parameter $\left< n_J - n_0 \right> / N$ measuring the macroscopic variable of interest. The abscissa measures the relative difference between kinase and phosphatase numbers. In this representation, all kinase to phosphatase proportions are compressed symmetrically between -1 and 1. As a result, the hyperbolic behavior of a single phosphorylation / dephosphorylation loop ($J=1$) without feedback is a straight line, the dotted diagonal. All other curves are for $J=2$, which is the minimal ultrasensitive case capable of bistability. The blue curve shows the ultrasensitivity of the chain without feedback. As $J\rightarrow\infty$, the slope of the sigmoidal approaches 2 at the midpoint $I=P$ (not shown). The blue curve already covers nearly 50\% of the ultrasensitivity attainable at $J=\infty$. Ultrasensitivity is enhanced by feedback (green curve), when approaching the threshold for bistability from below. The strength of feedback is reported in terms of $g=N/\sqrt{IP}$, where $N$ is the number of substrate molecules (see text for details). In the case of the green curve $g = 2.85$.  The slope at the midpoint is always $\infty$ at the critical point for the onset of bistability in the presence of feedback. Above the threshold (red curve; $g=5$) there is an intermediate range of $\left(I - P \right) / \left(I + P \right)$ with two stable solutions. The dots connecting the two branches indicate a sudden change in stationary phosphorylation state as one of the branches ceases to exist when $\left(I - P \right) / \left(I + P \right)$ passes outside the bistable range. This creates hysteresis.

\subsection*{Figure 3 - Phase diagrams.}

(a) Symmetric feedback circuit as in Fig.~\ref{fig:schemes}d. (b) Asymmetric feedback circuit (from the fully phosphorylated $J$-state) as in Fig.~\ref{fig:schemes}c. The coupling parameter $g = N / \sqrt{IP}$, the exogenous catalytic strength $P/I$, and the number of phosphoacceptor sites $J$ are being varied. Symbols are simulation results ($N=1000$) and solid lines are analytical results from the mean-field theory for $J =2,3,5$ (solid, dash, dot). The discrepancies arise from fluctuations, which the mean-field theory ignores. In the circuit with symmetric feedback, there is one critical coupling strength $g_c$ for each ratio of phosphatase to kinase numbers. (The phase diagram for symmetric feedback is itself symmetric about $0$ in $\log \left( I/P\right)$.  Only half of the diagram is shown in panel (a); the symmetric half is obtained by reflection about the vertical axis.) As $g$ is increased beyond $g_c$, the system exhibits two stable distributions of phosphorylation states peaked at the end states of the chain. Below the critical number, the target molecules are mostly unphosphorylated ($P/I>1$) and the system remains in this state as it becomes bistable. Within the bistable regime, the system could be prepared in the mostly phosphorylated state, in which it   persists as the number of target molecules is increased. Yet, as the number of target molecules is decreased, the mostly phosphorylated state loses stability abruptly and the system shifts to the mostly unphosphorylated state. This is a first-order phase transition in statistical mechanics. In contrast, for the symmetric feedback circuit {\it and\/} $P/I=1$ (the leftmost point on the abscissa in panel a), the transition is second-order, that is, continuous in the phosphorylation state. This is shown in Fig.~\ref{fig:meanfield} below. Panel (a) reveals that, once bistable, the symmetric circuit never loses bistability again as $N$ is further increased. In the language of statistical mechanics, only a lower critical coupling exists. Notice that for a fixed coupling $g$ and increasing $P/I$ the symmetric system loses bistability again, as illustrated in Fig.~\ref{fig:bistable}. Unlike the symmetric case, the asymmetric circuit exhibits a window of bistability (lower and upper critical couplings) in $N$. For a suitable $P/I$, an increase in the number of target molecules $N$ drives the system through a second threshold at which bistability disappears. If the system was mostly unphosphorylated, it now undergoes a (first-order) phase transition to the mostly phosphorylated state.

\subsection*{Figure 4 - $\xi(g)$ as obtained from inverting Eq.~\ref{eqn:g_as_sinh_full} for $J=2$.}

At a fixed ratio of kinase to phosphatase numbers, $\exp{\lambda}$, the abscissa measures the total number of substrate molecules $N$ in terms of $g$. The ordinate, $\xi=\log x$, expresses (at fixed $\lambda$ and $J$) the same information as the order parameter $\left< n_J - n_0 \right> / N$ by virtue of Eq.~\ref{eqn:ord_parameter_sinh}. The colors indentify different values of $\lambda$, indicated at the top on the right hand side. The negative sign means that there is more phosphatase than kinase in the system, and the only stable distribution below the critical point has $\left< n_J\right> < \left< n_0 \right>$. i.e. $x<1$. The blue curve for $\lambda=0$ depicts the second-order phase transition (also rendered in Fig.~\ref{fig:meanfield}, where it is compared against  simulations for different $J$). At $g_c=3$ the $x=1$ solution becomes unstable (dotted continuation), branching off continuously into two stable stationary states. Parameter asymmetry ($\lambda\ne 0$) leads to a discontinous change of the order parameter (first-order phase transition), which becomes apparent when the system is prepared in the upper branch and the control parameter $g$ is decreased below critical. At that point, the system jumps to the stationary state on the lower curve of the corresponding color. Again, dotted segments indicate the unstable solution.

\subsection*{Figure 5 - Order parameter $\left| \left< n_J - n_0 \right> \right| / N$ for the case $I=P$ in the symmetric circuit.}

The modulus $\left| \cdot\right|$ folds the two symmetric branches (corresponding to the two stationary state distributions) into one. The solid curves show the analytical results from the mean-field theory and the symbols represent simulations.  $J+1 = \left[ 5, 10, 100 \right]$ is shown in [blue, green, red] for the mean-field theory and as $\left[ \mbox{+}, \mbox{o}, \mbox{$\times$} \right ]$ for simulations. Target molecule numbers used in the simulations are, respectively, $N = [4000, 2000, 400]$.
    
\subsection*{Figure 6 - Fluctuations in the order parameter.}

The fits (solid lines) are leading order expansions about the mean-field solution obtained through the field-theory formalism (see text). The straight line has slope $1/g$. Symbols, $J+1$ values, and colors are as in Fig.~\ref{fig:meanfield}.

\subsection*{Figure 7 - Residence times $\tau$ in the switched state as a function of $J$, $N$, and the distance from the critical point $g-g_c$.}

In all cases the circuit has symmetric feedback and $I=P$.
(a): $\log \tau$ vs. $N$ for $J=2,3,4,5$ based on Monte-Carlo simulations. To meaningfully compare residence times, the coupling $g=N/\sqrt{IP}$ is kept constant by varying the external kinase and phosphatase numbers appropriately, $I=P=N/g$. To compare across $J$, the distance from the critical point, $g-g_c$, is held constant at 1 (recall that $g_c=(J+1)/(J-1)$). The green curve is for $J=2$ and $g = 4.0862$ (from setting $\xi = 1$ in Eq.~\ref{eqn:g_as_sinh}). This is the case calculated from first principles in \cite{smith:07}. The best fit to the slope of the green simulation data is $\log \tau \sim N*0.023$, which is very close to our theoretical prediction $\log \tau \sim N* 0.021$.
(b): $\log\tau$ vs. $g-g_c$ for different $J$s at $N=50$. Curves from bottom to top are for $J=2,3,4,5,6,7,8,9$.
  
\begin{comment}

%%%%%%%%%%%%%%%%%%%%%%%%%%%%%%%%%%%
%%                               %%
%% Tables                        %%
%%                               %%
%%%%%%%%%%%%%%%%%%%%%%%%%%%%%%%%%%%

%% Use of \listoftables is discouraged.
%%

\section*{Tables}
  \subsection*{Table 1 - Sample table title}
    Here is an example of a \emph{small} table in \LaTeX\ using  
    \verb|\tabular{...}|. This is where the description of the table 
    should go. \par \mbox{}
    \par
    \mbox{
      \begin{tabular}{|c|c|c|}
        \hline \multicolumn{3}{|c|}{My Table}\\ \hline
        A1 & B2  & C3 \\ \hline
        A2 & ... & .. \\ \hline
        A3 & ..  & .  \\ \hline
      \end{tabular}
      }
  \subsection*{Table 2 - Sample table title}
    Large tables are attached as separate files but should
    still be described here.

%%%%%%%%%%%%%%%%%%%%%%%%%%%%%%%%%%%
%%                               %%
%% Additional Files              %%
%%                               %%
%%%%%%%%%%%%%%%%%%%%%%%%%%%%%%%%%%%

\section*{Additional Files}
  \subsection*{Additional file 1 --- Sample additional file title}
    Additional file descriptions text (including details of how to
    view the file, if it is in a non-standard format or the file extension).  This might
    refer to a multi-page table or a figure.

  \subsection*{Additional file 2 --- Sample additional file title}
    Additional file descriptions text.

\end{comment}

\end{bmcformat}
\end{document}